\documentclass[twocolumn,prd,aps,showpacs]{revtex4}

\usepackage{amsmath}
\usepackage{verbatim}
\usepackage{graphicx}
\usepackage[usenames]{color}
\usepackage{psfrag}
\usepackage{hyperref}

\def\beq{\begin{equation}}
\def\eeq{\end{equation}}
\def\bea{\setlength\arraycolsep{1.4pt}\begin{eqnarray}}
\def\eea{\end{eqnarray}}
\def\bit{\begin{itemize}}
\def\eit{\end{itemize}}

\def\ie{{i.e.}}
\def\eg{{e.g.}}

\def\bra{\langle}
\def\ket{\rangle}

\def\alm{a_{\ell m}}

\def\Cl{C_\ell}

\begin{document}

\title{Observing the evolution of the CMB}

\author{Adam Moss} \email{adammoss@phas.ubc.ca}
\author{James P. Zibin} \email{zibin@phas.ubc.ca}
\author{Douglas Scott} \email{dscott@phas.ubc.ca}
\affiliation{Department of Physics \& Astronomy\\
University of British Columbia,
Vancouver, BC, V6T 1Z1  Canada}

\date{\today}

\begin{abstract}

Most cosmological parameters are expected to change significantly only on cosmological time scales, but given the large amount of information contained within the Cosmic Microwave Background (CMB) sky, we can expect that changes in the CMB should be observable on much shorter time scales.  Here we quantify this expectation, examining the detectability of the dominant effects on short time scales.  We find that an ideal future experiment with currently achievable sensitivity could detect the changing dipole due to our galactic motion in about 10 years, but that it would take around 4000 years to detect a change in the higher order multipoles.

\end{abstract}

\maketitle


{\em Introduction.\,} One of the most scientifically profitable endeavors of recent times has been the increasingly accurate measurement of temperature anisotropies in the Cosmic Microwave Background (CMB) (see~\cite{ss06}, and references within, for reviews). These measurements have been at the cornerstone of constructing a  standard model of cosmology, coined the $\Lambda$CDM model, which is seeded by near scale invariant adiabatic initial conditions created during inflation. 

The CMB sky is normally regarded as a ``snapshot'' of conditions on our last scattering surface (LSS) (together with some line-of-sight effects), with a finite information content which could be extracted by a sufficiently sensitive experiment.  However, in recent work~\cite{Zibin:2007mu}, we have discussed extensively the time evolution of the CMB in the context of a $\Lambda$CDM cosmology.  This raises the possibility that one might learn more about the Universe by observing the changing CMB sky, \ie\ by using the time domain, as is common with other astrophysical objects.

The main evolutionary properties of the CMB can be classified into five categories: (1) an overall cooling of the mean temperature; (2) a changing dipole due to our evolving velocity with respect to the CMB frame; (3) new structure from probing a growing LSS; (4) geometrical effects of projection as the distance to the LSS grows; and (5) effects due to evolving structure along the line of sight. For small time scales ($\lesssim 10^{6} $ years), the second of these effects dominates on large angular scales, and the third on small scales \cite{Zibin:2007mu}.  As a result, if one were to take the difference between the observed CMB sky now and in the near future, the resulting map would be dominated by small scale features around the anisotropy damping scale of  $\sim 10$ arcminutes.

These effects are to be distinguished from the behaviour of the CMB at very late times in a $\Lambda$CDM cosmology, which contains a future event horizon.  Somewhat akin to the stagnation a distant observer sees as objects approach the event horizon of a black hole, in the cosmological case an observer will see the pattern of CMB anisotropies ``freeze in'' (accompanied by ever increasing redshift) as the sphere of the LSS approaches a maximum radius on the de Sitter time scale (see, \eg, \cite{loeb02}).

Since the sensitivity of CMB experiments has improved dramatically over the years, in this letter we extend our treatment in Ref.~\cite{Zibin:2007mu}, and ask how long it would take to detect a change with current and future technology. This issue has also been briefly considered in Ref.~\cite{Lange:2007tx}, but here we take a more thorough approach to the detection statistics and also consider evolution of the CMB dipole.


{\em Detection criteria.\,} The  goal of this letter is to quantify when one could detect the temporal evolution of the CMB, so here we must first clarify this condition. A natural basis to expand the temperature fluctuations of the CMB are the spherical harmonics $Y_{\ell m} (\bf{\hat{n}})$, such that
$\delta T ({\bf \hat{n}}, \eta ) /T (\eta)= \sum_{\ell m} a_{\ell m} (\eta) Y_{\ell m} (\bf{\hat{n}})\,,$
where $\eta$ is the conformal time. The statistical properties governing the time evolution of the $a_{\ell m}$ coefficients are then encoded in the matrix (assuming isotropy)
\begin{equation}
\bf{C} = \left(\begin{matrix} C_{\ell}^{\eta \eta} & \Cl^{\eta\eta^{\prime}}  \cr  \Cl^{\eta \eta^{\prime}} & C_{\ell}^{\eta^{\prime} \eta^{\prime}} \end{matrix}\right)\,,
\end{equation}
where $\Cl^{\eta\eta'} \equiv {\rm Re}\,\bra\alm(\eta)a^\ast_{\ell m}(\eta')\ket_S$ is a correlation function.  Here, $\bra\dots\ket_S$ denotes the ensemble average over $a_{\ell m}$ (signal) realizations.  In particular, when $\eta = \eta'$ this correlation function reduces to the ordinary anisotropy power spectrum, $\Cl^{\eta\eta} = \Cl(\eta)$.

We find it convenient to define the power spectrum of the {\em  difference} between $a_{\ell m}$  modes at two times,  $D_{\ell}  = \langle  | \delta a_{\ell m}  |^2 \rangle_S$, where $\delta a_{\ell m} = a_{\ell m} (\eta^{\prime})-a_{\ell m}(\eta)$. Expanding this quantity, one finds $D_{\ell} = C_{\ell}(\eta) + C_{\ell}(\eta') - 2C_{\ell}^{\eta \eta^{\prime}}$. Since $D_{\ell}$ contains covariance information between the $a_{\ell m}$s at different times, it should be better  to measure this, rather than simply the change in power spectra, $\delta C_{\ell}= C_{\ell} (\eta^{\prime}) -C_{\ell} (\eta)$, if one were to attempt to detect a change.

Unfortunately, when making CMB sky maps to compare at different times, each of the $a_{\ell m}$ modes will be corrupted by noise, such that $a_{\ell m}^{\rm map}=a_{\ell m} + a_{\ell m}^{\rm noise}$. Furthermore, since we only sample a finite number of $a_{\ell m}$ modes from the ensemble, one can only provide estimators of the true variance.  If one constructs $C_{\ell}^{\rm map } = \sum_{m } |a_{\ell m}^{\rm map}|^2 /(2\ell+1)$, then an appropriate estimate of the variance is $C_{\ell}^{\rm est} = C_{\ell}^{\rm map }  - N_{\ell} $, where we assume the noise variance is also isotropic, i.e. $N_{\ell} = \langle |a_{\ell m}^{\rm noise}|^2  \rangle_N$.  Here $\langle \ldots \rangle_N$ refers to the average over the ensemble of noise realizations only.  At this point, however, we need to make the distinction between two situations involving  $C_{\ell}^{\rm est}$: 

(A) The averaging  is taken over all possible signal {\em and} noise realizations, and is denoted $\bra\dots\ket$. The ensemble average of the variance estimate is $\langle C_{\ell}^{\rm est} \rangle=C_{\ell}$, and so the estimate is unbiased. In standard parameter estimation analysis, one is often interested in the width of this distribution, i.e. $(\Delta C_{\ell})^2= \langle (C_{\ell}^{\rm est}-C_{\ell})^2\rangle$. In a full sky CMB experiment it can be shown that this quantity is given by~\cite{Knox:1995dq, Dodelson:2003ft} $ \label{eqn:delta_cl}
\Delta C_{\ell}=\sqrt{\frac{2}{2\ell+1}} (C_{\ell} + N_{\ell})\,.
$
Hence, even if $N_{\ell}$ is zero there is still sampling uncertainty of the underlying signal variance. This is commonly referred to as ``cosmic variance''. 

(B) In the second situation the signal component is fixed and we average {\em only} over the ensemble of noise realizations.  Hence, we are interested solely in the uncertainty of the measurement of a given signal due to noise corruption, {\em not} the uncertainty in its underlying value. To emphasize the fact that we now regard the signal as fixed, we denote the realization of the  $a_{\ell m}$ component that we are dealt by $\bar{a}_{\ell m}$. If we construct the variance of this component by $\bar{C}_{\ell} = \sum_{m } |\bar{a}_{\ell m}|^2 /(2\ell+1)$, then we now find that $\langle C_{\ell}^{\rm est} \rangle_N=\bar{C}_{\ell}$.  Furthermore, the width of this distribution, $(\Delta C_{\ell}^N)^2= \langle (C_{\ell}^{\rm est}-\bar{C}_{\ell})^2\rangle_N$, is given by
\beq \label{eqn:delta_cl_n}
\Delta C^N_{\ell}=\sqrt{\frac{2 N_{\ell}}{2\ell+1} \left(2 \bar{C}_{\ell} + N_{\ell} \right)}\,.
\eeq
In this case, the uncertainty in measurement vanishes when the experimental noise is zero, as one would expect. 

Applying these situations to the time evolution of the CMB, one can therefore ask two different questions:  
has the realization of the sky that we have been {\em dealt} changed within experimental noise uncertainty (corresponding to case B), or has our realization changed to such an extent that we can conclude that the ensemble averaged spectrum has changed (corresponding to case A)? Clearly, the former criterion is easier to meet than the latter.

\begin{figure}
\centering
\mbox{\resizebox{0.35\textwidth}{!}{\includegraphics{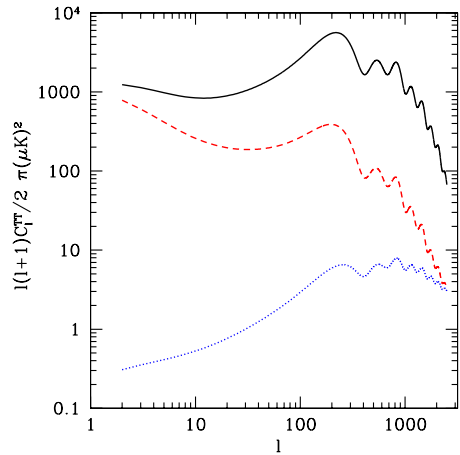}}}
\caption{\label{fig:cl_noise}  Primary CMB temperature anisotropies (solid curve) compared to the error in $C_{\ell}$. The dashed curve shows $\Delta C_{\ell}$, and the dotted curve $\Delta C_{\ell}^N$. We assume Planck noise levels.}
\end{figure}

To illustrate the size of the errors given by $\Delta C_{\ell}$ and  $\Delta C_{\ell}^N$, in Fig.~\ref{fig:cl_noise} we plot these quantities along with the primary CMB temperature anisotropies for an instrument with $N_{\ell}=1\times 10^{-4}\, \mu {\rm K}^2/T^2$.   Such a noise level will be obtainable by the upcoming Planck experiment~\cite{bluebook}. On large scales, $\Delta C_{\ell}$ is comparable to the primary anisotropies due to sample variance. However, even though we may be uncertain about the underlying variance, we can in principle measure $\bar{C}_{\ell}$ with high precision: $\Delta C_{\ell}^N$ is considerably lower than $\Delta C_{\ell}$.

Equivalent expressions apply in pixel space---in this case the temperature at each pixel $i$ of the CMB sky map contains a signal and noise contribution: $(\delta T_i/T)^{\rm map} = (\delta T_i /T)+ (\delta T_{i}/T)^{\rm noise}$. The expected signal variance over the ensemble for each pixel is $C_S=\langle (\delta T_{i}/T)^2 \rangle_S=\sum_{\ell} (2\ell+1)C_{\ell}/(4 \pi)$ (i.e. the total power in the map), and the noise variance $C_N=\langle (\delta T_{i}^{\rm noise}/T)^2 \rangle_N =\sigma_{\rm pix}^2$. If one constructs $C^{\rm map}=\sum_{i} (\delta T_{i}^{\rm map}/T)^2/N_{\rm pix}$, where $N_{\rm pix}$   is the total number of pixels in the map, then an appropriate estimate of the variance is $C_S^{\rm est}=C^{\rm map}-C_N$. If one defines $\bar{C}_S=\sum_{i} (\delta T_{i}^{}/T)^2/N_{\rm pix}$, then averaging over the ensemble of noise realizations gives $\langle C_S^{\rm est} \rangle_N= \bar{C}_S$. The equivalent error on the signal variance to that given above becomes $\Delta C_S^N = \sqrt{2 C_N\left(2 \bar{C}_S + C_N \right)/N_{\rm pix}}$.  Analogous quantities are defined for the difference map power~$D$.

Using these expressions we can estimate the time-scale for detectability. However,  it will first be instructive to consider a special case before considering higher order CMB modes. 


{\em Time varying dipole.\,} The CMB dipole ($\ell = 1$) is determined by the Doppler shift of the approximately isotropic CMB radiation field due to our observer motion with respect to its rest frame.  The magnitude of the dipole  variance is much higher than modes with $\ell \ge 2$, and the evolution has contributions on much shorter time scales.  Indeed, the time variation of the Earth around the Sun has already been well measured~\cite{Kogut:1993ag}, and is used to calibrate CMB experiments, so here we consider larger scale solar motions.

For simplicity, we decompose our observer velocity into two components: $\bf{v} = \bf{v}_{\rm Sun-GC} + \bf{v}_{\rm GC-CMB} $. The first component  is due to our solar motion around the Galactic centre (GC), and the second is due to motion of the GC with respect to the local group and other large scale structures. We assume our galactic motion is confined to the plane, with a tangential velocity of $222$ ${\rm km \,s^{-1}}$ at a radius of 8.5 kpc, giving a galactic rotation period $T_g$ of $2.35 \times 10^8$ years. We assume the $\bf{v}_{\rm GC-CMB} $ component remains constant on galactic time-scales.

The best estimate of the measured $a_{1 m}$ components come from the WMAP satellite~\cite{Hinshaw:2006ia}, and are given in fractional temperature units by $\left( {a}_{1 -1} ,  \, {a}_{1 0}, \, {a}_{1 1} \right) = \left( v_y, \, v_z, \, v_x \right) = \left( -816.0, \, 919.3, \, -87.8 \right) \times 10^{-6} \,,$ with around $0.5\%$ calibration uncertainty.  The predicted evolution of the dipole components due to galactic motion is then  $\left( {a}_{1 -1}  ,  \, {a}_{1 0}  , \, {a}_{1 1}  \right) = 740.5 \, \left(  \cos x, \, 0, \,  \sin x  \right) \times 10^{-6}  +  \left(  -1556.5, \, 919.3 , \,  - 87.8 \right) \times 10^{-6}  \,,$ where $x=2\pi t/T_g$ and $t$ is the local observer time.

For small time periods $t \ll T_g$, we can compute the evolution in the dipole variance ${C_{1}} =\sum_{m} |{a}_{1 m}|^2 /3$ to be ${C}_1 (t) \approx {C}_1 (t_0) -1.16 \times 10^{-15} \,t/{\rm years}$, where ${C}_1 (t_0) = 5.062 \times 10^{-7} $. The evolution in the difference power is $D_{1} = 1.30  \times 10^{-22} \,(t/{\rm years})^2$. Although $\delta C_{1}$ is larger than $D_{1}$  over small time-scales, the larger noise error means it is  a less sensitive statistic, as we show below.

The simplest estimate of the time-scale for detectability is as follows. We assume that we have a full sky data-set with {\it no noise\/} today, and use the predicted evolution of the $a_{1 m}$ coefficients to generate another noiseless sky at some time in the future. Assuming the pixel noise is uncorrelated, one constructs the hypothesis that the two skies are different within the expected noise. So we form
$\chi^2 = \sum_{i}^{N_{\rm pix}} \left[T_i/T(t) - T_i/T (t_0)\right]^2/(2C_N)$\,,
where the factor 2 in the noise arises since we are differencing maps.  Under the null hypothesis, the expectation for this quantity is zero and the standard deviation is $\sqrt{2N_{\rm pix}}$, so one can evaluate the significance level $\alpha$  that the two skies are different.  Note that for the dipole, we only have noise averaging, as the signal is deterministic, unlike the $\ell \ge 2$ modes.

The noise contributes on all angular scales (not just the $\ell=1$ mode),  and may swamp the signal variance of the difference map, $D_S$.  In the noise dominated regime, the simple $\chi^2$ statistic is equivalent to the signal-to-noise (S/N) ratio $\alpha=D_S/ \Delta D_S^N$. In an experiment such as Planck, the large number of pixels ($\approx 4 \times 10^6$) will beat down the noise level, and one finds $\alpha \approx 2$ after $t=2 \times 10^4$ years, even though the map is noise dominated.

Much of this noise, however, resides at $\ell \ge 2$. If one performs a harmonic transform on the noisy map, one can recover a higher S/N by isolating the $\ell=1$ mode. In this case, the uncertainty in the signal variance given the noise is $\Delta D_S^N=3 \Delta D_{1}^N/(4\pi)$. At $t=2 \times 10^4$ years, by filtering small scale noise, the detection significance increases to $\alpha \approx 40$.

Finally, one could also compare the change in power $\delta C_{1}$. The uncertainty in $C_{1}$ due to noise at each time is $\Delta C_{1}^N$.  Since $C_{1} (t) \approx C_{1} (t_0)$, $\Delta C_{1}^N (t) \approx \Delta C_{1}^N (t_0) = \Delta C_{1}^N$, and the S/N is $\alpha=|\delta C_{1}|/(2\Delta C_{1}^N)$. After $t=2 \times 10^4$ years,  an experiment with Planck noise gives $\alpha \approx 4$. Therefore, the loss of time covariance information has led to a reduction in statistical significance over the filtered difference map method. If one instead assumed the signal was stochastic,  and used $\Delta C_{1}$ instead of $\Delta C_{1}^N$, one would obtain a tiny significance, as one is estimating the probability that two uncorrelated samples have a different underlying ensemble variance.

\begin{figure}
\centering
\mbox{\resizebox{0.35\textwidth}{!}{\includegraphics{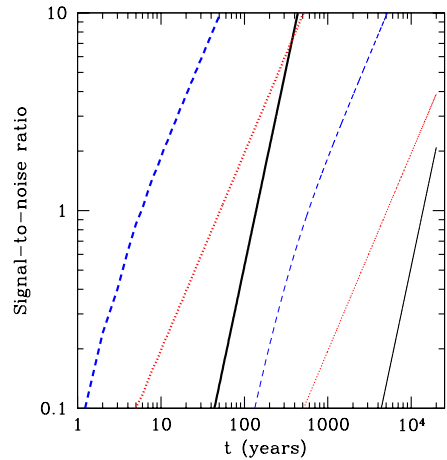}}}
\caption{\label{fig:dip_prob}  Ideal case detection signal-to-noise for observing a change in the CMB dipole. Thin lines show an instrument with Planck sensitivity, and thick lines an instrument 100 times as sensitive. Solid curves show the significance using the noise dominated difference map, and the dashed curves after removing small scale noise. Dotted curves are for the dipole magnitude alone, and hence contain no covariance information.}
\end{figure}

In this idealized case, we plot the detection S/N in Fig.~\ref{fig:dip_prob} for Planck noise levels, along with an instrument 100 times as sensitive (this is current generation technology~\cite{Bock:2006yf}). In principle, such an instrument could observe the evolving galactic dipole on a $\sim 10$ year time-scale.

An important systematic in any such experiment would be calibration uncertainty, the accuracy required being many orders of magnitude better than what has been achieved so far. There are two contributions to dipole evolution---a change in amplitude due to changing speed relative to the CMB frame, and a change in orientation due to rotation. By normalizing $C_{1} (t) = C_{1} (t_0)$, hopefully eliminating calibration issues, the only contribution is due to rotation. Due to our position in the galaxy (our velocity is mainly against the bulk velocity flow),  the rotation component actually dominates currently, and amplitude normalization only leads to a  degradation in S/N of  $< 0.5\%$.  Another more insidious effect is time-varying foreground emission. As well as the motion of Galactic  clouds, the observed foreground pattern will change on the same time-scale as the CMB dipole.  It is difficult to place a degradation factor on these effects, although one would hope that multi-wavelength observations would minimize them. 


{\em Higher order anisotropies.\,} In contrast with the dipole, the evolution of modes with $\ell  \ge 2$ depends only on the set of cosmological parameters. For small time intervals $t$ ($\lesssim 10^{6} $ years), the spectral shape of the change in variance $\delta C_{\ell}$ and difference power $D_{\ell}$ are fixed, with the amplitudes scaling $\propto t$ and $\propto t^2$ respectively \cite{Zibin:2007mu}. 
The shape of $D_{\ell}$ is relatively featureless, being approximately constant up to the anisotropy damping scale. On scales smaller than this  ($\ell > 1000$) the power falls off rapidly.

\begin{figure}
\centering
\mbox{\resizebox{0.35\textwidth}{!}{\includegraphics{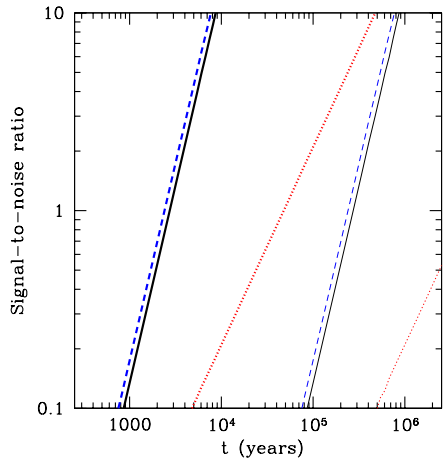}}}
\caption{\label{fig:cmb_prob}  Ideal case detection signal-to-noise for observing a change in the higher order ($\ell \ge 2$) modes. Labels are the same as in Fig.~\ref{fig:dip_prob}.}
\end{figure}

Truncating the spectrum at $\ell = 2000$ (the approximate resolution of Planck), the total expected difference power is given by $D_S = 3.11\times 10^{-23}\, (t/19.8 \, {\rm years})^2$.
The prefactor in this expression is the dipole signal variance, $3 D_{1} /(4\pi)$,
so it would take about 20 times longer than for the dipole to obtain the same S/N $\alpha = D_S/\Delta D_S^N$. This is borne out in Fig.~\ref{fig:cmb_prob}, with around $4000$ years required to detect a change in the $\ell \ge 2$ modes at the $\alpha=2$ level using current detector technology. 

In the case of the dipole, we were able to reduce this nominal value significantly by removing small scale noise.  We follow a similar procedure here, computing  the expected S/N at each  $\ell$ mode, $\alpha_{\ell}=D_{\ell}/\Delta D_{\ell}^N$, and discarding modes with low significance. In order to compute the combined significance, we first rank the modes in terms of $\alpha_{\ell}$. Applying a weight to the signal at each $\ell$ mode, given by $w_{\ell}=(\Delta D_{\ell}^N)^{-2}$, the combined signal and noise components are then given by $\sum_{\ell} w_{l} D_{l}/ \sum_{\ell} w_{l}$ and $(\sum_{\ell } w_{\ell})^{-1/2}$ respectively. By  iterating through the modes, starting from high $\alpha_{\ell}$, we compute the combined significance until the maximal S/N is obtained. Typically, this procedure retains modes in the range $20 \lesssim \ell \lesssim 500$, so the limiting factor in any such measurement is instrument sensitivity rather than resolution. Unfortunately, since the signal contributes on a range of scales, similar improvements to the dipole are not possible here, and the increase in S/N is small, as shown in  Fig.~\ref{fig:cmb_prob}. Our results for $\ell \ge 2$ are consistent with Ref.~\cite{Lange:2007tx}, who use $\Delta D_{\ell}$ as the noise, without the cosmic variance term. Since the signal at each $\ell$ mode is noise dominated, $\Delta D_{\ell} \approx \Delta D_{\ell}^{N}$ in this regime.

Following the same method using the expected change in variance, $\delta C_{\ell}$, such that the S/N at each $\ell$ mode is $\alpha_{\ell}=\delta C_{\ell}/(2 \Delta C_{\ell}^N)$, the time for detection increases by around a factor of ten before one could detect a statistical change. 

Where does the stochastic nature of the signal for the $\ell \ge 2$ modes enter into this treatment? For small time intervals, the evolution of anisotropies is primarily driven by probing new structure at the surface of last scattering. The spectrum $D_{\ell}$ of this structure has a corresponding sample variance, equal to $2 D_{\ell}^{2}/(2\ell+1) $ at each $\ell $ mode. In these estimates, we have assumed that the future signal we would measure is at the peak of the likelihood function (equal to $D_{\ell}$). The uncertainty in the exact signal we would measure induces a corresponding spread on the S/N, but we found this to be negligible after performing realizations of the future sky. 


{\em Conclusions.\,}  The time domain provides critical information in many branches of astrophysics, even cosmology (distant supernovae and active galactic nuclei being obvious examples). However, the Universe as a whole is evolving slowly and so one expects this evolution to be difficult to observe, as is the case with the Sandage-Loeb test \cite{sandage62,loeb98} for the increase in redshift of fixed sources.

In this letter we answered the question of how long one would have to wait to see new information from the gold-mine of cosmological data that is the CMB.  The large quantity of data available means that we could observe a change on time scales much shorter than cosmological.  By considering optimal estimates for differences in observed skies we estimate that, using currently available detector technology, the dipole can be observed to change due to our Galactic motion in about a decade, whereas it would take thousands of years to detect even a statistical change in the higher order multipoles. 

{\em Acknowledgments.\,} This research was supported by the Natural Sciences and Engineering Research Council of Canada.  We thank Phil Gregory, Mark Halpern and Jasper Wall for useful discussions.


\end{document}